# Improved 2D Intelligent Driver Model simulating synchronized flow and evolution concavity in traffic flow

Junfang Tian[1]*, Rui Jiang[2], Geng Li[1]*, Martin Treiber[3], Chenqiang Zhu[1], Bin Jia[2]

[1]*Institute of Systems Engineering, College of Management and Economics, Tianjin University, Tianjin 300072, China*

[2]*MOE Key Laboratory for Urban Transportation Complex Systems Theory and Technology, Beijing Jiaotong University, Beijing 100044, China*

[3]*Technische Universität Dresden, Institute for Transport & Economics, Würzburger Str. 35, D-01062 Dresden, Germany*

This paper firstly show that 2 Dimensional Intelligent Driver Model (Jiang et al., PloS one, 9(4), e94351, 2014) is not able to replicate the synchronized traffic flow. Then we propose an improved model by considering the difference between the driving behaviors at high speeds and that at low speeds. Simulations show that the improved model can reproduce the phase transition from synchronized flow to wide moving jams, the spatiotemporal patterns of traffic flow induced by traffic bottleneck, and the evolution concavity of traffic oscillations (i.e. the standard deviation of the velocities of vehicles increases in a concave/linear way along the platoon). Validating results show that the empirical time series of traffic speed obtained from Floating Car Data can be well simulated as well.

**Key words:** driving behavior; synchronized traffic flow; evolution concavity

## 1. Introduction

Traffic flow exhibits many complex fascinating phenomena, such as the phantom jam, the traffic breakdown and related capacity drop, the widely scattered data of congested flow on the flow-density plane and multifarious macroscopic patterns (see Brackstone and McDonald, 1999; Chowdhury, 2000; Helbing, 2001; Nagatani, 2002; Brackstone et al., 2009; Treiber and Kesting, 2013; Kerner, 2004, 2009, 2013; Saifuzzaman and Zheng, 2014; Zheng, 2014). In order to understand the traffic flow dynamics, various traffic flow models have been proposed, including car following models (such as Kerner and Klenov, 2002; Zhang and Kim, 2005; Kerner and Klenov, 2006; Kesting and Treiber, 2008; Li et al. 2013; Chen et al. 2014), cellular automata models (such as Nagel and Schreckenberg, 1992; Kerner et al., 2002; Jiang et al. 2003; Hafstein et al. 2004; Kerner et al. 2011; Tian et al., 2015a) and hydrodynamic

---

*Corresponding author.

Email address: jftian@tju.edu.cn, ligeng345@126.com

models (such as Lighthill and Whitham, 1955; Kerner, 1993; Helbing et al. 2001; Li and Zhang, 2001; Jiang et al. 2002; Zhang, 1998, 1999, 2003; Tang et al, 2014a and 2014b) and so on. Various experiments and empirical investigations are conducted to clarify driving behaviors (such as Fairclough et al. 1997; Jiang et al., 2014; Tian et al., 2015b; Yeung and Wong, 2014; Chung 2015).

Car Following Models (CFMs) aim to describe the longitudinal interactions between vehicles on the road based on the idea that drivers control their vehicles to react to the stimulus from preceding vehicles. In CFMs, the acceleration of a vehicle is usually expressed as a function of the space gap, relative velocity and acceleration information. General Motor (GM) models are one of the earliest and most well-known CFMs (Chandler et al., 1958; Gazis et al. 1961; Edie, 1961), which considered drivers' reaction delay and assumed that the acceleration of the following vehicle is proportional to its relative velocity to the preceding vehicle. The proportionality coefficient is assumed to depend on its velocity and its space gap to the preceding vehicle. Due to its simplicity, GMs have been extensively studied (Brackstone and McDonald, 1999; Saifuzzaman and Zheng, 2014).

Newell (1961) proposed a different kind of model hypothesizing that drivers adapt their speeds, with a reaction delay time, to the optimal speed, which is a nonlinear function of the space gap of their vehicles. In contrast to the GM models, this model is complete in the sense that it can also describe free traffic flow (Treiber and Kesting, 2013). Three decades later, Bando et al. (1995) presented the renowned Optimal Velocity Model (OVM) that can be regarded as a variant of Newell model via Taylor's expansion. OVM could reproduce the properties of traffic flow, such as the instability of traffic flow, the evolution of traffic congestion, and the formation of stop-and-go waves. Helbing and Tilch (1998) carried out the calibration of OVM with respect to the empirical data. Their simulation results show that OVM exhibits too high acceleration and unrealistic deceleration. Therefore, the Generalized Force Model (GFM) was proposed taking the negative velocity difference into account. Jiang et al. (2001) found that GFM underestimates the kinematic wave speed. They put forward the Full Velocity Difference Model (FVDM) by the consideration of both the positive and negative velocity difference. Zhao et al. (2005) argued that accidents occur in GFM and FVDM under urgent braking cases. They established the Full Velocity and Acceleration Difference Model (FVADM) with the incorporation of acceleration difference. There are many other extensions of OVM and FVDM, considering various aspects such as the reaction delay (Davis, 2003), the asymmetric acceleration and deceleration (Gong et al., 2008), the next-nearest-neighbor interaction (Sawada, 2001; Tordeux and Seyfried, 2014), and the interactions of arbitrary number of vehicles ahead (Lenz et al. 1999; Hasebe et al. 2003).

In 2002, Newell (2002) established another simple and popular model with the assumption that the trajectory of a vehicle in congested traffic on a homogeneous freeway is identical to that of its leading vehicle with some space and time shifts. Due to the limitation of the Newell model for the simulation of the characteristics of traffic oscillations, Laval and Leclercq (2010), Chen et al. (2012a, 2012b) and Laval et al. (2014) have made different extensions by considering driver's timid and aggressive behaviors and stochastic characteristics.

On the other hand, Kometani and Kometani (1959) proposed a collision avoidance model based on the assumption that the following vehicle aims to pursue a safe following gap. However, in their model, the vehicle's acceleration process has not been described. Gipps (1981) proposed a model supposing that the following vehicle's velocity is determined in the way to ensure safety even if the preceding vehicle brakes abruptly. Nonetheless, Treiber and Kesting (2013) pointed out that Gipps model does not differentiate between comfortable and maximum deceleration. If the desired deceleration in Gipps model is considered as the maximum deceleration, then the model is accident free but every braking maneuver is performed uncomfortably with full brakes; if the desired deceleration is regarded as the comfortable deceleration, then Gipps model possibly produces accidents in heterogeneous and/or multi-lane traffic provided preceding vehicles' deceleration exceeds the desired deceleration. Treiber et al. (2000) presented the Intelligent Driver Model (IDM), which assumes that in the equilibrium state the space gap to the leading vehicle is determined by the minimum gap and desired time gap to the leading vehicle. In other situations an 'intelligent' braking strategy is imposed to keep smooth comfortable accelerations and decelerations.

Except for GMs, all above mentioned models assumed the existence of a unique relationship between space gap and velocity in the steady state of traffic flow. The traffic flow in these models can be classified into free flow and congested traffic flow. Thus they are named as two-phase models by Kerner (2013)[†]. In contrast, in Kerner's Three-Phase Traffic theory (KTPT), congested traffic flow is further classified into synchronized flow and jam (Kerner, 2004, 2009). The steady state of synchronized traffic flow is assumed to occupy a two-dimensional region in the flow-density plane. Kerner argued that phase transition involved in two-phase models is a transition from free flow to jams (F→J). However, in real traffic and in KTPT, one usually observes the phase transitions from Free flow to Synchronized flow (F→S transition) and from Synchronized flow to Wide moving jams (S→J transition). Thus, the transition from free flow to jams corresponds to a F→S→J process. In an open road section with an isolated bottleneck, two types of spatiotemporal traffic patterns can be observed: the general pattern and the synchronized flow pattern, which can be further classified into the widening synchronized pattern (WSP), the localized synchronized pattern (LSP), and the moving synchronized pattern (MSP).

Recently, Jiang et al. (2014) carried out the controlled car following experiments concerning a platoon of 25 passenger cars on a 3.2-km-long open road section. The leading vehicle was asked to move with different constant speed. The evolution concavity of oscillation growth has been identified, i.e., the standard deviation of the velocities of vehicles increases in a concave/linear way along the platoon. Later, the evolution concavity of oscillation growth has also been reported in the empirical data (Tian et al., 2015b). Via comparisons with the simulation results of the two-phase car following models, Jiang et al. (2014) concluded that the nature of car following runs against those models such as OVM, FVDM and IDM. They found that by removing the fundamental assumption of unique relationship between space gap and velocity in two-phase models and allowing the traffic state to span a two-dimensional region in velocity-spacing plane, the growth pattern of disturbances has changed and becomes qualitatively in accordance with the observations. In particular, the IDM with two-dimensional region (2D-IDM) can fit the experimental results quantitatively well.

However, unfortunately, as shown in section 2, the 2D-IDM fails to simulate the synchronized traffic flow.

---

[†] Although GMs do not assume the existence of a unique relationship between space gap and velocity in the steady state of traffic flow, they are also classified as two-phase models by Kerner since the GMs cannot reproduce the F→S→J process, either.

Motivated by this fact, this paper proposes an improved 2D-IDM by considering the defensive driving behavior in high speed traffic. The improved model is shown to be able to, and to our knowledge, it is the first car-following model that can, simultaneously reproduce the empirical spatiotemporal patterns, phase transitions from synchronized flow to wide moving jams, and evolution concavity in traffic flow.

The paper is organized as follows. Section 2 analyzes the 2D-IDM, shows its deficiency and presents the improved model. Section 3 presents simulation results of the improved model. Section 4 verifies the improved model, using the empirical floating car data. Section 5 concludes this paper.

## 2. The Model

*2.1. The 2D-IDM*

The two main equations of the 2D-IDM (Jiang et al., 2014) are formally the same as that of the IDM. The acceleration $a_n$ of vehicle $n$ is given as a function of its speed $v_n$, the spacing $d_n(t) = x_{n+1}(t) - x_n(t) - L_{veh}$ to the leading vehicle $n+1$, where $L_{veh}$ is the length of the vehicle, and the relative speed $\Delta v_n(t) = v_{n+1}(t) - v_n(t)$ by

$$a_n(t) = a_{max}\left(1 - \left(\frac{v_n(t)}{v_{max}}\right)^4 - \left(\frac{d_{n,desired}(t)}{d_n(t)}\right)^2\right) \quad (1)$$

$$d_{n,desired}(t) = \max\left(v_n(t)T(t) - \frac{v_n(t)\Delta v_n(t)}{2\sqrt{a_{max}b}}, 0\right) + d_0 \quad (2)$$

where $v_{max}$ is the maximum velocity, $a_{max}$ is the maximum acceleration, $b$ is the comfortable deceleration, $d_0$ is the jam gap. In contrast to the IDM where $T$ is a constant parameter, the desired time gap $T(t)$ of the 2D-IDM is a stochastic quantity changing its value in each simulation step $\Delta t = 0.1s$ as follows:

$$T(t + \Delta t) = \begin{cases} T_1 + rT_2 & \text{with probability } p, \\ T(t) & \text{otherwise.} \end{cases} \quad (3)$$

where $r$ is uniformly distributed random number between 0 and 1. The three new parameters $p$, $T_1$, and $T_2$ replace the IDM time-gap parameter $T$. Specifically, the parameters $T_1$, and $T_2$ indicating the range of the time gap variations gives rise to two-dimensional flow-density data in congested steady states.

Fig.1 shows the flow-density and speed-density diagrams of 2D-IDM, which is formed from the initial homogeneous distribution. The free flow branch ($K<K_c$) and the wide moving jam branch ($K>K_c$) can be identified, while the synchronized flow branch is not reproduced (see Fig.4 for the comparison). In the free flow branch, vehicles move with the maximum speed with some fluctuations (see Fig.2(a)). In the wide moving jam branch, the large localized disturbances emerge from the free flow and propagate in downstream direction. Then, the speeds of some vehicles fall to zero and the wide moving jams appear, which propagate in upstream direction (see Fig.2(b)). Therefore, the transition reproduced by 2D-IDM is the unrealistic F→J transition.

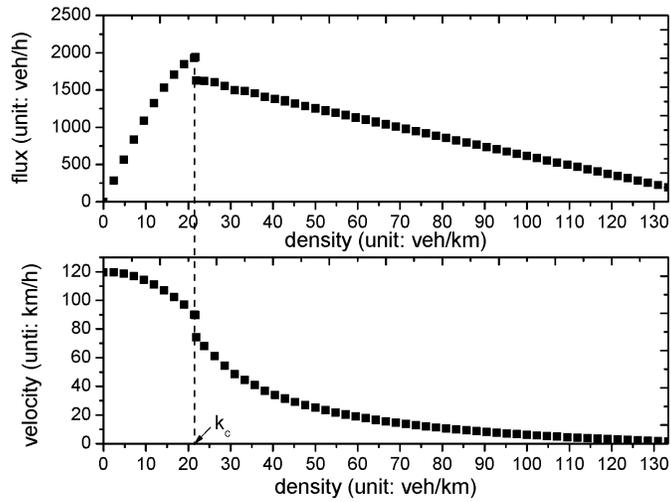

**Fig. 1.** Flow-density and speed-density diagrams of 2D-IDM. In the simulation, the parameters are set as: $v_{max}$=120$km/h$, $a_{max}$=0.73$m/s^2$, $b$=1.67$m/s^2$, $d_0$=2$m$, $T_1$=0.5$s$, $T_2$=1.9$s$, $p$=0.015, and $L_{veh}$ = 5$m$.

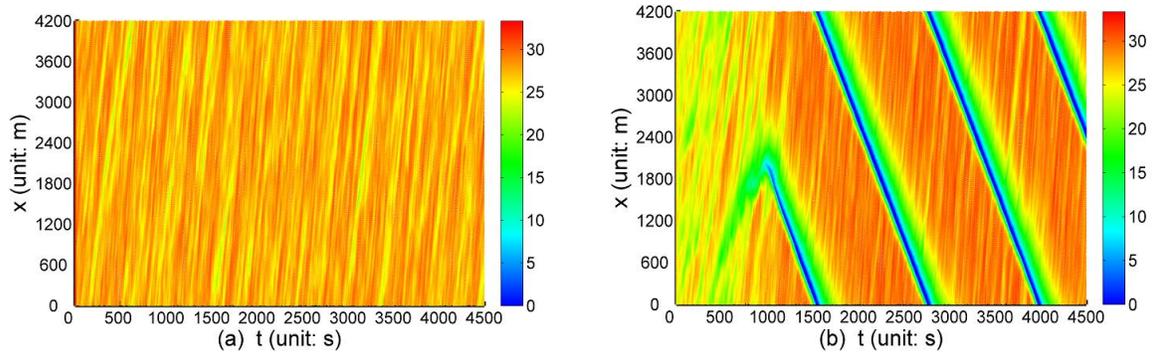

**Fig. 2.** The spatiotemporal diagrams of velocities (Unit: m/s) in 2D-IDM on the circular road. In (a, b) the density $k$=19, 22 *veh/km*, respectively.

## 2.2. The Improved 2D-IDM

We argue that the failure to reproduce the synchronized traffic flow in 2D-IDM can be attributed to that the difference between driving behaviors at high speeds and that at low speeds has not been considered. In order to keep safety, drivers tend to be more defensive in the high speed state, in particular under the circumstance that the actual space gap is smaller than the desired space gap. Based on the fact, an improved model based on the improved IDM (Treiber and Kesting, 2013) is proposed as follows:

If $d_{n,\text{desired}}(t) \leq d_n(t)$

$$a_n(t) = a_{\max}\left(1-\left(\frac{v_n(t)}{v_{\max}}\right)^4\right)\left(1-\left(\frac{d_{n,\text{desired}}(t)}{d_n(t)}\right)^2\right)$$

else

$$\begin{cases} \text{If } v_n(t) \leq v_c \\ \quad a_n(t) = a_{\max}\left(1-\left(\frac{d_{n,\text{desired}}(t)}{d_n(t)}\right)^2\right) \\ \text{else} \\ \quad a_n(t) = \min\left(a_{\max}\left(1-\left(\frac{d_{n,\text{desired}}(t)}{d_n(t)}\right)^2\right), -b\right) \end{cases} \quad (4)$$

where $v_c$ is the critical velocity. $T(t)$ is still the desired time gap but redefined as:

$$T(t+\Delta t) = \begin{cases} T_1 + rT_2 & \text{if } r_1 < p_1 \text{ and } v_n(t) \leq v_c, \\ T_3 + rT_4 & \text{if } r_1 < p_2 \text{ and } v_n(t) > v_c, \\ T(t) & \text{otherwise.} \end{cases} \quad (5)$$

where $r$ and $r_1$ are still two independent random numbers between 0 and 1. These revisions have considered the situation that when vehicles are at high speeds ($v_n(t) > v_c$), drivers tend to become more defensive. The improved 2D-IDM is abbreviated as 2D-IIDM in the text below.

## 3. Simulation analysis

Firstly, we simulate traffic flow on a circular road and on an open road with a bottleneck as one normally tests a traffic flow model. We study whether the synchronized traffic flow and the widening synchronized pattern (WSP) can be simulated or not. Next, car-following behaviors have been simulated to test whether 2D-IIDM can preserve the advantage of 2D-IDM to reproduce the experimental evolution feature of disturbances and the spatiotemporal patterns as revealed in the car-following experiment (Jiang et al., 2014). In the simulation, the parameters are set as: $v_{\max}$=120$km/h$, $v_c$=50.4$km/h$, $a_{\max}$=0.8$m/s^2$, $b$=1.5$m/s^2$, $d_0$=2.0$m$, $T_1$=0.5$s$, $T_2$=1.9$s$, $T_3$=0.9$s$, $T_4$=1.5$s$, $p_1$= $p_2$=0.015, and $L_{\text{veh}} = 5m$.

*3.1 Circular road*

The following two initial configurations are used in the simulations: 1) all vehicles are homogeneously distributed on the road; 2) all vehicles are distributed in a megajam. Fig.3 shows the flow-density and velocity-density diagrams of 2D-IIDM. In the branch that the density is smaller than $K_1$, there is only free flow on the road (Fig.4(a)).

Compared with that of 2D-IDM (Fig.1), there are two branches in the density region $K_1 < K < K_2$. In this region, the upper branch exhibits the synchronized traffic flow (Fig.4(b)), which initiates from the initial homogeneous distribution; the lower branch is the coexistence state of free flow and wide moving jam (Fig.4(c)), which comes from the initial megajam. When the density is larger than $K_2$, the synchronized flow is unstable, and wide moving jams will appear finally. Fig.4(d) shows the spontaneous S→J transition, which is consistent with the observed empirical phenomenon. Therefore, the synchronized traffic flow, the metastable states and the S→J transition are successfully depicted in the 2D-IIDM.

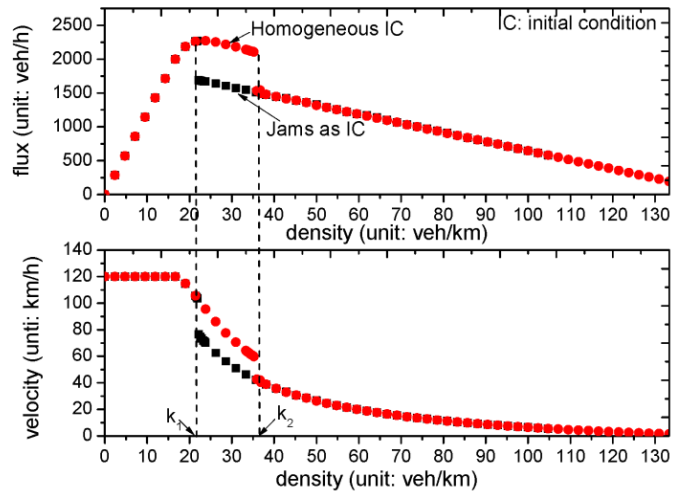

**Fig. 3.** Flow-density and speed-density diagrams of 2D-IIDM.

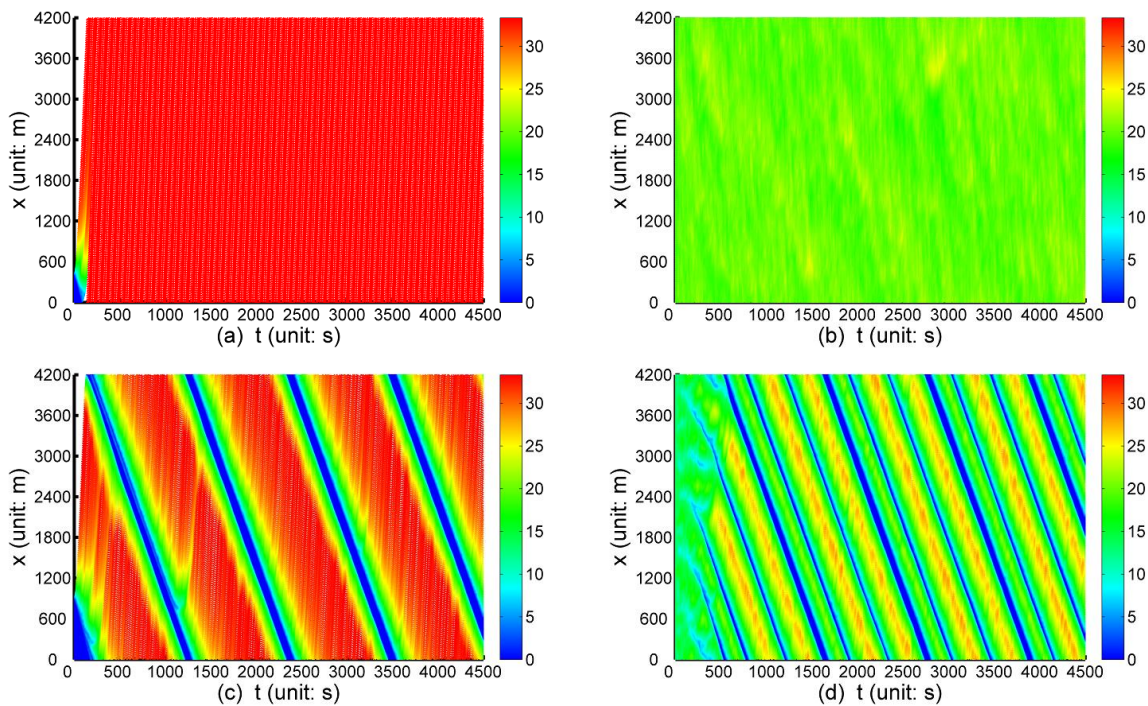

**Fig. 4.** The spatiotemporal diagrams of velocities (Unit: *m/s*) in 2D-IIDM on the circular road. From (a) to (d), the density *k* is 19, 31, 31, 43*veh/km*. In (a, c) the traffic starts from a megajam. In (b, d) the traffic starts from a homogenous distribution.

*3.2 Open road with a rubbernecking bottleneck*

Road bottlenecks include flow-conserving ones (like gradients, a local decrease of the speed limit, local road narrowings inducing a locally increased time gap and so on) or rubbernecking bottleneck, and flow-non-conserving ones, such as on-ramps, off-ramps. In this paper, we consider a rubbernecking bottleneck on an open road as in Chen et al. (2012a). It should be noted that any type of flow-conserving bottleneck as well as on-ramp bottlenecks will have the same effect.

The rubbernecking zone is located at [0.9$L_{road}$, 0.9$L_{road}$+300]*m*, where the road length $L_{road}$ = 700*km*. When vehicles enter this zone, at each simulation time step, they have a probability $\gamma$ to rubberneck which will cause their speeds to decrease instantaneously by $\varphi$%. Rubbernecking only can occur at most once in this zone. Initially, the road section is assumed to be filled with vehicles uniformly distributed with density $\rho$ and their velocities are set to $v_{max}$. For the leading vehicle, it will be removed when it goes beyond $L_{road}$. The second car becomes new leading car and it moves freely. At the road section entrance, no vehicle is inserted. This simulation setup is used because high flow rate cannot be achieved by using usual boundary setup.

Fig.5 presents the simulation results in the road segment [0.9$L_{road}$-4000, 0.9$L_{road}$+2000]*m*. Fig.5(a) shows the spatiotemporal features of the widening synchronized flow (WSP), where only the upstream front propagates upstream and no wide moving jams appear. Fig.5(b) gives the local synchronized pattern (LSP) where the synchronized flow is only localized in the vicinity of the bottleneck region. Fig.5(c) shows the General Pattern (GP) that wide moving jams continuously emerge in synchronized traffic flow. Fig.5(d) shows the Dissolving General Pattern (DGP) that only one wide moving jam emerges in synchronized traffic flow and there is the LSP near the bottleneck region. Therefore, most of the empirical findings, especially the WSP, are successfully simulated by 2D-IIDM.

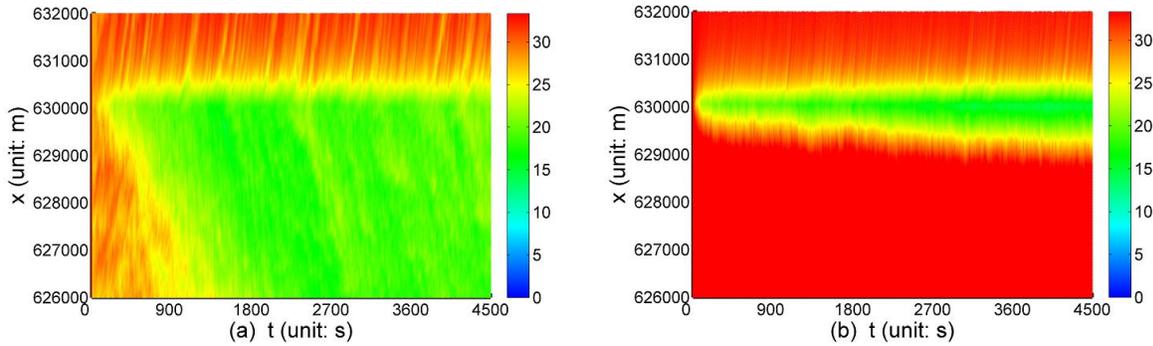

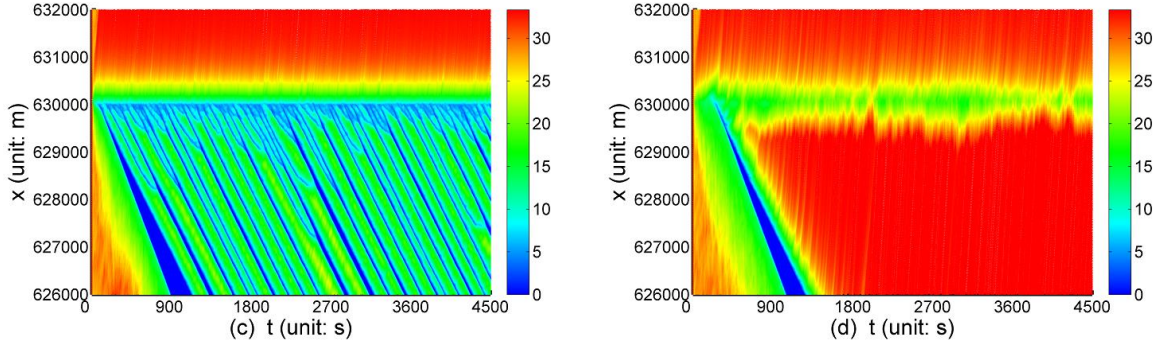

**Fig. 5.** The spatiotemporal diagrams of 2D-IIDM on an open road with a rubbernecking. (a) $\rho=22.8veh/km$, $\gamma=0.025$, $\varphi= 2\%$ (WSP), (b) $\rho=16veh/km$, $\gamma=0.04$, $\varphi= 7\%$ (LSP), (c) $\rho=22.8veh/km$, $\gamma=0.03$, $\varphi= 30\%$ (GP) and (d) $\rho=22.8veh/km$, $\gamma=0.02$, $\varphi= 10\%$.

*3.3 Car-following experiment simulation*

The platoon of 25 vehicles are investigated. All vehicles are initially distributed in a megajam. The leading vehicle accelerates and moves exactly the same as in the experiment. Fig.6 shows that the growth rate of the disturbances is in line with the experimental data. Fig.7 shows that under various velocity $v_l$, the formation and evolution of oscillations closely resemble that of Jiang's experiments. Therefore, the evolution concavity of oscillation growth is well simulated by 2D-IIDM.

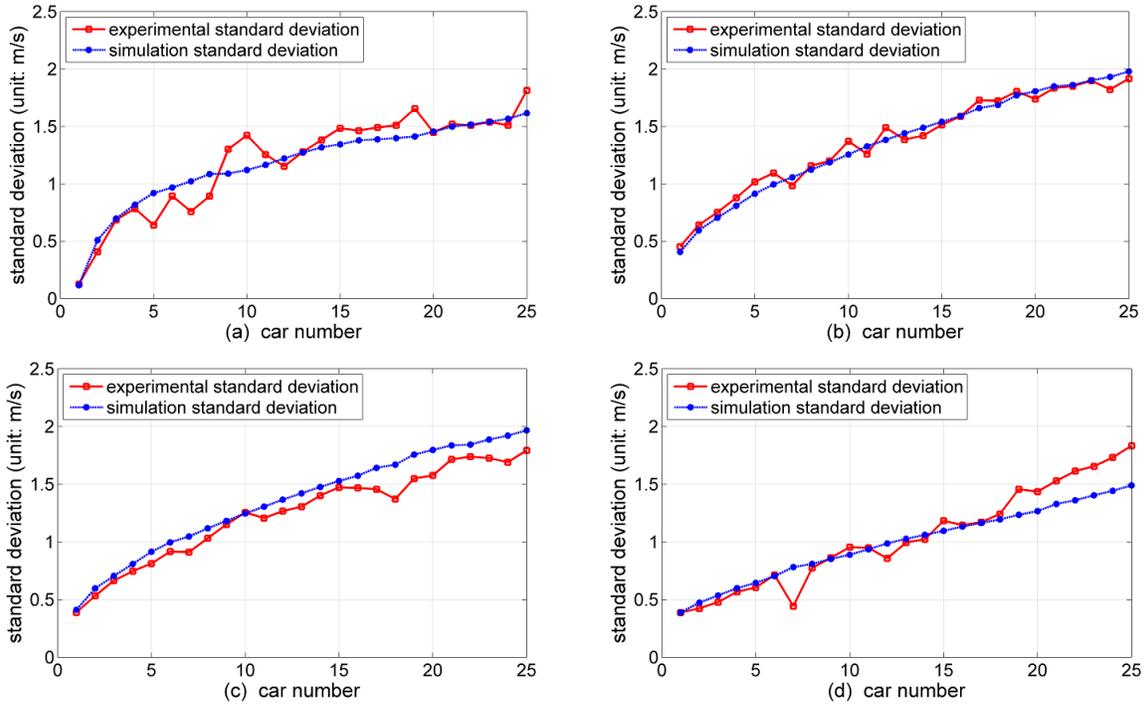

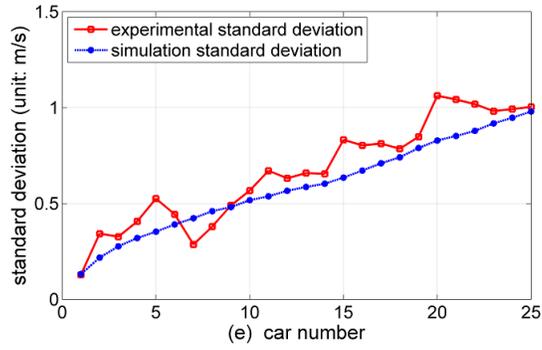

**Fig. 6.** Comparison of the simulation results and experiment results of the standard deviation of the velocities of the cars. The car number 1 is the leading car. $v_l$ =50, 40, 30, 15, 7$km/h$ in (a)-(e), respectively.

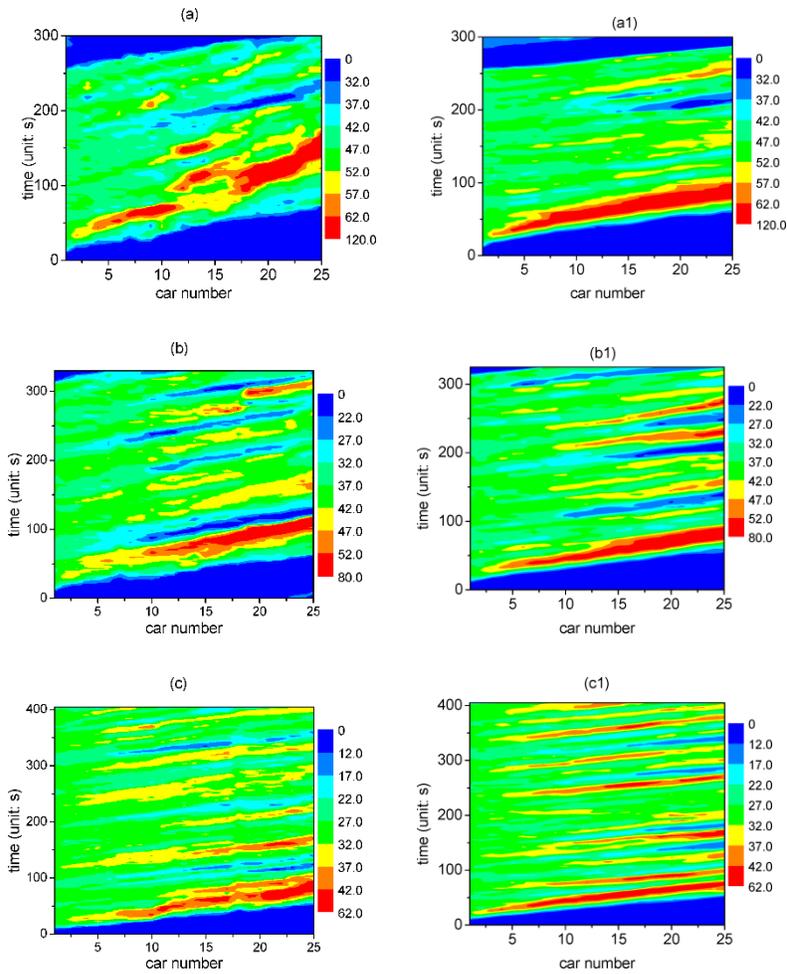

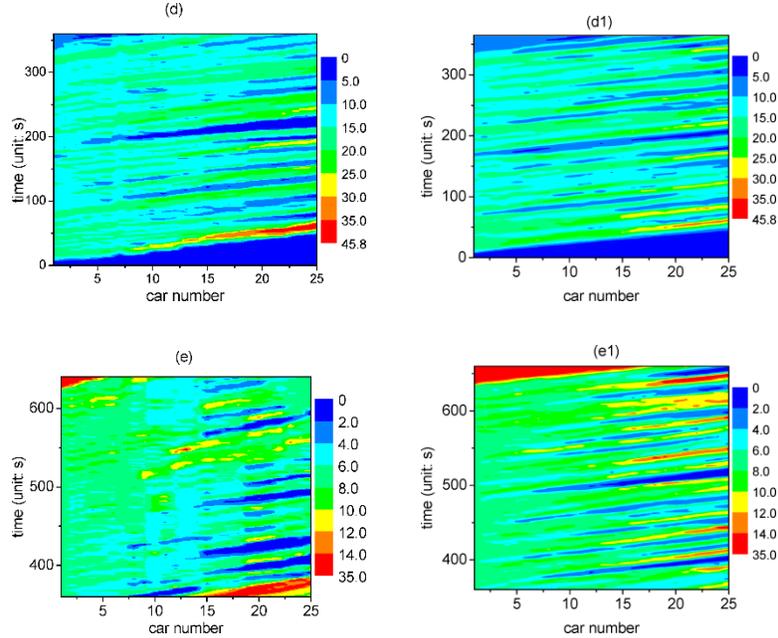

**Fig. 7.** The spatiotemporal patterns of the platoon traffic. The car speed is shown with different colors (unit: *km/h*) as function of time and car number. The left and right Panels show the experimental results and the simulation results of 2D-IIDM. In (a, a1) - (e, e1), the leading vehicle is required to move with $v_l$=50, 40, 30, 15, 7*km/h* respectively. In the simulation, the velocity of the leading car is set the same as in the experiment.

## 4. Floating Car Data Validation

Three Floating Car Data (FCD) sets (Kesting and Treiber, 2008) are applied to test the performance of 2D-IIDM, which are provided by Robert Bosch GmbH (DLR, 2007) and recorded during an afternoon peak period on a fairly straight one-lane road in Stuttgart, Germany. A car equipped with a radar sensor in front provides the relative speed and gap to the car ahead. The durations of the measurements are 400*s*, 250*s* and 300*s*, respectively. The data are recorded with a frequency of 10*Hz*, i.e. with a time increment of 0.1*s*. The first two sets are limited to the low speed situations below approximately 6*m/s*, corresponding to the gaps smaller than 12*m* most of the time. In set 3, the speed varies in the range of 0*m/s* and 18*m/s*, which contains a jump of the gap to the leader from approximately 20*m* to 40*m* at a time of about 144*s* because of a passive lane change of the leader.

During the simulation, we calibrated the model parameters and the results show that the maximum speed of 2D-IIDM a needs to be adjusted to 60*km/h* and other parameters the same as that in Section 3. Since 2D-IIDM is a stochastic model, the probability distribution of the following vehicles was estimated through repeated simulations of the lead vehicle problems (LVPs, Laval et al., 2014). Fig.8 shows the validation results. One can see that all FCD profiles almost fall into the 90%-probability bands, which means that the movements of the following vehicles can be predicted by 2D-IIDM very well.

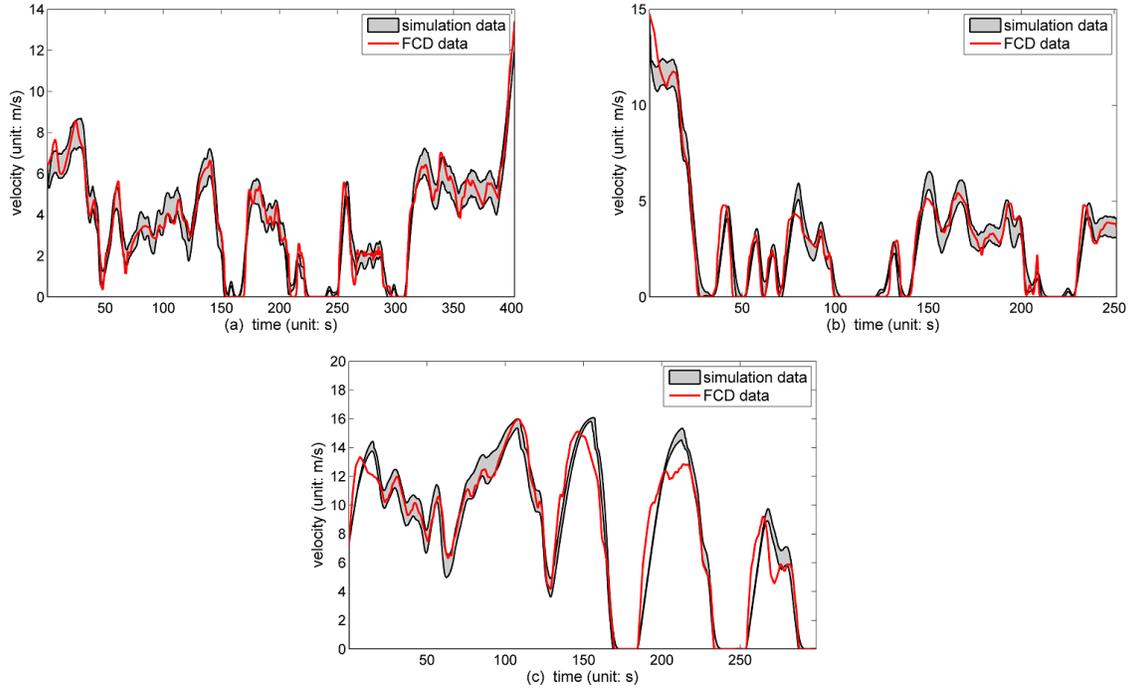

**Fig. 8.** Comparison of simulated and FCD trajectories.

## 5. Conclusion

Car Following Models (CFMs) aim to describe the longitudinal interactions between vehicles on the road. Traditional two-phase CFMs usually assume that there is a unique relationship between space gap and velocity in the steady state. However, Kerner (2013) pointed out that two-phase CFMs cannot simulate the empirical F→S→J transitions. Jiang et al. (2014) also reported that these two-phase CFMs have exhibited the evolution convexity of the initial growth of oscillation, which runs against the evolution concavity in the experimental and empirical findings (Jiang et al., 2014; Tian et al. 2015b). The 2D-IDM has been proposed, which can simulate the evolution concavity quantitatively well. However, unfortunately, we showed in this paper that the 2D-IDM fails to simulate the synchronized traffic flow.

This paper proposes an improved 2D-IDM, by considering the defensive driving behavior of vehicles at the high speed states. Simulations show that the 2D-IIDM can reproduce most of identified empirical phenomena. Specifically, simulations on a circular road show that 2D-IIDM can simulate the metastable state, and phase transition from synchronized flow to wide moving jams. Simulations on an open road with a rubbernecking bottleneck demonstrate that most of the spatiotemporal patterns of traffic flow can be well described, especially the widening synchronized pattern. Simulations on an open road with a moving bottleneck (realized by a slow moving leading vehicle in a platoon) illustrate that the evolution concavity of traffic oscillations can be well reproduced. Furthermore, validating results show that the speed time series from the Floating Car Data can be well depicted. To our knowledge, it is the first car-following model that can simultaneously reproduce so many empirical findings.

Nevertheless, the 2D-IIDM is still not able to reproduce the first-order F→S transition. In our future work, more efforts are needed to furthermore improve the model. Moreover, experimental and/or empirical trajectory data need to be collected to identify the F→S transition.

## Acknowledgements

JFT was supported by the National Natural Science Foundation of China (Grant No. 71401120, 71431005). RJ was supported by the Natural Science Foundation of China (Grant Nos. 11422221 and 71371175). GL was supported by the Tianjin social science planning project (Grant Nos. TJGL13-015). BJ was supported by the National Natural Science Foundation of China (Grant No. 71222101).